\documentclass{iau-JDSS}
\usepackage{graphicx}
\usepackage{natbib}

\title[SpS16. Cosmic Fullerenes] 
{Fullerenes in circumstellar and interstellar environments}

\author[Cami]   
{Jan Cami}
\affiliation{Department of Physics \& Astronomy, The University of
  Western Ontario, \\ London ON N6A 3K7, Canada ({\tt jcami@uwo.ca})}

\begin{document}
\maketitle
\begin{abstract}
In recent years, the fullerene species C$_{60}$ (and to a lesser
extent also C$_{70}$) has been reported in the mid-IR spectra of
various astronomical objects. Cosmic fullerenes form in the
circumstellar material of evolved stars, and survive in the
interstellar medium (ISM). It is not entirely clear how they form or
what their excitation mechanism is. 

\end{abstract}

\firstsection 
\section{Fullerenes in astrophysical environments}
Fullerenes (such as the buckminsterfullerene C$_{60}$) are large
carbonaceous molecules in the shape of a hollow sphere of
ellipsoid. They are very stable, and thus it was suggested early on
that they could also form in space and be abundant and widespread in
the Universe \citep{1985Natur.318..162K}.
Astronomical searches for the electronic bands of neutral C$_{60}$
were unsuccessful though \citep[for an overview, see][]{Herbig:C60};
and the detection of two diffuse interstellar bands near the predicted
wavelengths of C$_{60}^{+}$ awaits confirmation from a gas-phase
laboratory spectrum \citep{Foing:C60_1}.
C$_{60}$ also has 4 IR active vibrational modes at 7.0, 8.5, 17.4 and
18.9 $\mu$m. Dedicated searches for these bands did not result in a
detection either \citep{1995AJ....109.2096C,Moutou:C60}. 

Recently, we reported the detection of the IR active modes of C$_{60}$
and C$_{70}$ in the Spitzer-IRS spectrum of the young, low-excitation
planetary nebula (PN) Tc~1 \citep{Cami:C60-Science}. Since then,
fullerenes have been found in many more PNe
\citep{Garcia-Hernandez:PN,Garcia-Hernandez:MC}, a proto-PN
\citep{ZhangKwok:proto-PNC60}, a few R~Cor~Bor stars
\citep{Garcia-Hernandez:RcrB,2011AJ....142...54C} and even in O-rich
binary post-AGB stars \citep{Gielen:C60p-AGB}. In addition, fullerenes
have turned up in interstellar environments
\citep{2010ApJ...722L..54S,2011MNRAS.410.1320R,2012ApJ...753..168B}
and in young stellar objects \citep{Roberts:C60} as well. From these
detections it is clear that fullerenes are formed in the circumstellar
environments of evolved stars. They then either survive in the ISM
(possibly incorporated into dust grains), or form there when
conditions are right.

\section{The fullerene excitation mechanism}
To explain the IR emission of cosmic fullerenes, two mechanisms have
been considered that offer quite different predictions about the
relative band strengths \citep[for a detailed comparison,
  see][]{Jero:C60excitation}. Thermal C$_{60}$ emission models show
large variations in the relative strength of all bands as a function
of temperature; for $T\le 300$~K, the 7.0 and 8.5 $\mu$m bands are
very weak compared to the 17.4 and 18.9 $\mu$m bands. For fluorescence
on the other hand, the band strengths only depend on the average
absorbed photon energy; in that case, the 17.4/18.9 $\mu$m band ratio
is roughly constant (for reasonable photon energies) while the 7.0 and
8.5 $\mu$m bands should be fairly strong.

Observationally, the 7.0 and 8.5 $\mu$m bands are often very weak or
even undetectable, while there are considerable variations in the
17.4/18.9 $\mu$m band ratio; this is more easily explained by thermal
models than by fluorescence models. However, these variations could
also be the consequence of contamination by PAH emission. In the three
known uncontaminated fullerene-rich PNe on the other hand, the 7.0
$\mu$m band is far {\em too strong} to be explained by even
fluorescence from C$_{60}$ alone. As pointed out to the careful reader
by \citet{Jero:C60excitation}, the 7.0 $\mu$m emission in those
sources includes a significant contribution from C$_{70}$, provided at
least that the emission is due to fluorescence. For one object (Tc~1),
fluorescence is further supported by the observation that the C$_{60}$
emission peaks at large distances ($\sim8000$~AU) from the central
star. If fluorescence is also the excitation mechanism for the other
astronomical sources where fullerenes have been detected, then the
weak 7.0 and 8.5 $\mu$m bands indicate that the fullerene emission is
not due to isolated, free C$_{60}$ molecules in the gas-phase; there
might be contributions from other species as well and/or the emission
may be due to fullerene clusters or nanocrystals.

\section{The formation of cosmic fullerenes}
Several routes have been proposed to explain the formation of
fullerenes in astrophysical environments. Densities in circumstellar
and interstellar environments are too low for bottom-up fullerene
formation on reasonable timescales
\citep{Micelotta:arophatics}. Fullerenes could form from the
processing of PAHs \citep{2012PNAS..109..401B}, but this requires
fine-tuned initial conditions.
A promising route starts from {\em arophatics} -- large clusters of
aromatic rings with aliphatic and olefinic bridges that originate from
a:C-H grains \citep{Micelotta:arophatics}. UV irradiation first
dehydrogenates and aromatizes such structures; subsequent C$_{2}$
ejection then shrinks down the resulting cages to C$_{60}$. Further
shrinking is inhibited by a high energy barrier. The spectral imprint
of the parent a:C-H grains in the IR spectra of fullerene-rich PNe
offers some observational support for this mechanism
\citep{Jero:C60excitation}.



\end{document}